\documentclass[12pt,fleqn]{article}
\usepackage{psfig}
\newcommand{\beq}{\begin{equation}}
\newcommand{\eeq}{\end{equation}}
\newcommand{\beqn}{\begin{eqnarray}}
\newcommand{\eeqn}{\end{eqnarray}}
\newcommand{\bea}[1]{\beq\begin{array}{#1}}
\newcommand{\eea}{\end{array}\eeq}
\newcommand{\summ}[2]{\sum\limits_{#1}^{#2}}
\newcommand{\eq}[1]{(\ref{#1})}

\newcommand{\diff}{\partial}

\newcommand{\NPPS}[3]{{\it Nucl. Phys. Proc. Suppl. }{\bf #1} (#2) #3}
\newcommand{\PL}[3]{{\it Phys. Lett. }{\bf #1} (#2) #3}

\newcommand{\PR}[3]{{\it Phys. Rev. }{\bf #1} (#2) #3}

\newcommand{\CMP}[3]{{\it Comm. Math. Phys. }{\bf #1} (#2) #3}

\newcommand{\PTP}[3]{{\it Prog. Theor. Phys. }{\bf #1} (#2) #3}
\newcommand{\PTPS}[3]{{\it Prog. Theor. Phys. Suppl. }{\bf #1} (#2) #3}
\begin{document}
\title{
~\vspace{-50mm}
\begin{flushright}   {\large ITEP-TH-43/99 \\ KANAZAWA 99-22}  \end{flushright}
\vspace{15mm}
{\bf The lattice $SU(2)$ confining string \\
    as an Abrikosov vortex}}
\author{
F.~V.~Gubarev$^a$,
E.--M.~Ilgenfritz$^b$,
M.~I.~Polikarpov$^a$,
\\
T.~Suzuki$^b$
\\[5mm]
$^{a}${\small ITEP, B.~Cheremushkinskaya 25, Moscow 117259, Russia}
\\
$^b${\small Institute for Theoretical Physics, Kanazawa University,} \\
{\small Kanazawa 920-1192, Japan}
}
\date{}

\maketitle

\begin{abstract}
Numerical data \cite{Bali-all} for the $SU(2)$ confining string in the
maximal abelian projection are analysed. The distribution of the electric
flux and monopole currents are perfectly described by the classical
equations of motion for the dual Abelian Higgs model. The mass of the vector
boson is equal to the mass of the monopole (Higgs particle) within
numerical errors. The classical energy per unit length of the Abrikosov
vortex reproduces 94\% of the full non-Abelian string tension.
\end{abstract}

\vspace{10mm}

{\bf 1.}  The monopole confinement mechanism \cite{general} in lattice
gluodynamics is confirmed by many lattice numerical calculations (see
reviews \cite{reviews} and references therein). The simplest effective
dynamical picture corresponding to this mechanism is the dual Abelian Higgs
model (AHM) \cite{maedan} with the Lagrangian:

\beq\label{AHMlagr}
{\cal L}_{AHM}=
\frac{1}{4 g^2} F^2_{\mu\nu}(B) + \frac{1}{2} |(\diff_\mu - i B_\mu)\Phi|^2 +
\lambda (|\Phi|^2-\eta^2)^2   \, \, ,
\eeq
where $B_\mu$ is the dual gauge field interacting with the monopole field
$\Phi$.  Electrically charged ``quarks'' can be introduced \cite{maedan}
using the Zwanziger formalism \cite{Zwa71}. The confinement of electric charges
is due to the dual Abrikosov string and exists already at the classical
level. The first fact in favor of an effective Lagrangian like
\eq{AHMlagr} was the numerical proof of the existence of the monopole
condensate in the confining phase of lattice gluodynamics \cite{su2cond}.
The further numerical investigation allowed to derive an effective monopole
Lagrangian \cite{monlag} which is in agreement with \eq{AHMlagr}. There are
various attempts to determine the parameters of the effective dual theory
\eq{AHMlagr} from phenomenological analysis and from calculations on the
lattice (e.g.  \cite{maedan,baker,finite}). The results can be
expressed through the gauge boson mass $m_V = g\eta$ and Higgs (monopole)
mass $m_H = 2\sqrt{2\lambda\eta^2}$. In most cases, the analyses have led
to the result $m_V \approx m_H$. This is the so-called Bogomolnyi
limit \cite{Bogom} of the AHM, where several exact results can be
obtained. In particular the classical Abrikosov string tension is:

\beq \label{Bogomten}
\sqrt{\sigma}  =  \sqrt{{\pi\over g^2} m^2}\, ,\, \,\, m = m_H =
m_V \, .
\eeq

Of course the simple model \eq{AHMlagr} cannot describe all properties of
QCD, and it is important to find the region of applicability of this
effective theory. An interesting example was found
by Bali, Schlichter and Schilling \cite{Bali-all} who have
presented results on the profile of the flux tube between heavy quarks
in $SU(2)$ lattice gauge theory. Working within the maximal Abelian
projection on a $32^4$ lattice at $\beta=2.5115$, they have studied the
correlators of the Abelian electric field $\vec{E}$ and of the density of
Abelian monopole currents $\vec{k}$ with Abelian Wilson
loops. Such correlators, divided by the expectation values of the Wilson
loops themselves, describe the electric field and monopole currents induced
by the presence of a static chromoelectric string. For Wilson loop in the
$zt$ plane only the $z$ component $E_z(\rho)$ of $\vec{E}$ and the 
azimuthal component
$k_\theta(\rho)$ of $\vec{k}$ were found to be non-vanishing
(here $\rho=\sqrt{x^2+y^2}$ is the distance from the center of the flux tube
and the azimuthal angle $\theta$ is defined as usual $\tan \theta=y/x$).
The authors of \cite{Bali-all} fit the
numerical data using specific parameterization of $E_z(\rho)$ and
$k_{\theta}(\rho)$ which respect the boundary conditions for
the continuum dual Abrikosov string.

{\bf 2.}
Below we continue the analytical analysis of the numerical data for the
profile of the confining string. We will avoid to use specific ans\"atze.
Instead, we work directly with the Ginzburg--Landau equations, fitting the
parameters of the dual AHM in such a way that the dual Abrikosov string
reproduces the numerical data for $E_z(\rho)$. Our experiments revealed that
the usual $|\Phi|^4$ form of the Higgs field potential \eq{AHMlagr} is
sufficient. The inclusion of higher order term $c_6|\Phi|^6$ does not change
the results and gives $c_6=0$. In the cylindrical coordinate system
$t,z,\rho, \,\, \rho^2=x_a x_a \, \, (a=1,2)$, for the unitary gauge
$\Im m~\Phi=0$, the static axially symmetric and infinitely long ANO vortex
can be parameterized by two functions $b(\rho)$ and $f(\rho)$:

\beq
B_a =
\varepsilon_{ab} \frac{x_b}{\rho} \left({1\over\rho}-b(\rho)\right)\, ,
\, \, \Phi = \eta f(\rho) \, , \,\, B_0=B_3=0 \, .
\eeq
The energy per unit length ({\it i.e.} the string tension $\sigma$) can be
easily derived from \eq{AHMlagr}:

\beq\label{tension}
\sigma=
\frac{2\pi}{g^2} \int\limits^{\infty}_{0} \rho d\rho \left\{
\frac{1}{2} ({1\over\rho}[\rho b]')^2 +
\frac{1}{2} m^2_V (f')^2 +
\frac{1}{2} m^2_V f^2 ({1\over\rho}- b)^2 +
\frac{1}{8}m^2_V m^2_H (f^2-1)^2
\right\}
\eeq
The variational principle gives rise to the following
equations of motion :

\beq\label{1} {1\over\rho}\left[ \rho
b'\right]' - {1\over\rho^2}\;b + m^2_V f^2 \left({1\over\rho}-b\right) =0
\, ,
\eeq
\beq\label{2} {1\over\rho} \left[ \rho f'\right]' = f
\left({1\over\rho}-b\right)^2 + \frac{1}{2} m^2_H f (f^2-1) \, ,
\eeq
which  should be supplied with the boundary conditions
\beq
b(0)= f(0) = 0\, , \,\, \lim\limits_{\rho\to\infty} b(\rho)= {1\over\rho} \,
,\, \, \lim\limits_{\rho\to\infty} f(\rho)= 1\, .
\eeq

Note that \eq{1} is nothing but the dual Ampere law,
$\vec{k}=\mbox{curl}\vec{E}$, in cylindrical coordinates. We have numerically
solved the classical equations (\ref{1},\ref{2}) fitting the
data for $E_z(\rho)$ of Ref.  \cite{Bali-all}. The
fitting curves are plotted in Fig.~1. The numerical values of the
parameters of the fit are the following :

\bea{ll} g/2\pi & =  0.9274 \pm 0.0066 \, , \\
m_V & =  0.5733 \pm  0.0383 = (1.3123 \pm 0.0771) \mbox{~GeV} \, ,
\label{param-values} \\
m_H & =  0.5945 \pm  0.0063 = (1.3614 \pm  0.0143) \mbox{~GeV}  \, .
\eea
The vector and
scalar masses are equal to each other within the numerical errors. The
physical meaning of this equality deserves a separate discussion although,
as we mentioned, several models predict it.  \mbox{}From a practical point
of view we can estimate the string tension using eq.  \eq{Bogomten}:

\beq
\sqrt{\sigma} \; = \; 0.1744 \pm 0.0231 \; \approx
\; ( 400.1 \pm 53.0 ) \mbox{~MeV} \; \approx \; 0.91
\; \sqrt\sigma_{SU(2)} \, \, .
\eeq
Here and in \eq{param-values} the dimensionless quantities correspond to
the lattice spacing $a=1$, to get the dimensional quantities
we use the lattice results \cite{Bali-all} corresponding to $\beta =
2.5115$: $a = 0.086 \mbox{~fm}$, $\sqrt\sigma_{SU(2)} = 440 \mbox{~MeV}$.

{\bf 3.}
One sees from Fig.~1 that, in spite of the good approximation to
$E_z(\rho)$, the magnetic current distribution is not perfectly well
reproduced by our model in the region $1<\rho<3$. Since the Ampere law
\eq{1} is an integral part of the dual AHM, we conclude that either the dual
description is invalid in this region or the numerical data are not reliable
here due to finite lattice spacing effects.  Ideally, to check the validity
of dual AHM one should perform the same lattice measurements for larger
values of $\beta$ when lattice configurations are closer to the continuum.
Such an enterprise is certainly worthwhile although is impossible
practically at the moment.  Instead, we try to estimate the effect of
hypercubic
discretization by repeating our procedure on a coarse lattice similar to the
one used in Ref. \cite{Bali-all}.  The discretized action corresponding to
(\ref{AHMlagr}) is

\beq\label{AHM-lattice-action}
S= {1\over {4 g^2}} \summ{x,\mu\nu}{} F^2_{x,\mu\nu} +
{1\over 2} \summ{x,\mu}{}    | \Phi_x - e^{i B_{x,\mu}}\;\Phi_{x+\mu}|^2
+ \lambda \summ{x}{} (|\Phi_x|^2-\eta^2)^2 \, \, ,
\eeq
where $F_{x,\mu\nu} = B_{x,\mu} + B_{x+\hat{\mu},\nu} - B_{x+\hat{\nu},\mu}
- B_{x,\nu}$, and $a = 1$. Since we consider an infinitely long ANO vortex
the problem becomes essentially two-dimensional. In
the unitary gauge $\Re e\Phi_x = \eta f_x,\,\, \Im m \Phi_x = 0$, the string
tension is

\beq\label{sigma-lattice} \sigma= {1\over {2 g^2}} \left\{ \summ{x}{}
(F_{x,12})^2 + m^2_V \summ{x,\mu=1,2}{} | f_x - e^{i
B_{x,\mu}}\;f_{x+\mu}|^2 + {1\over 4} m^2_H m^2_V \summ{x}{}(f_x^2-1)^2
\right\}  , \eeq

Contrary to the continuum considerations above there are no apriori
ans\"atze for the ANO string solution on the lattice. Nevertheless, it is
known how to introduce a single vortex to the system \cite{MIP-vortex}: the
recipe is to make a shift $F_{x,12} \to F_{x,12} + 2\pi \delta_{x,x_0}$ in
the action (\ref{sigma-lattice}). Of course, periodic boundary conditions
are not allowed anymore. We have solved the corresponding equations of
motion on the square lattices $32^2 \div 70^2$ and with free boundary
conditions. We did not find any finite volume effects. Note that a  
$32^2$ lattice in our analysis corresponds to the $32^4$ lattice of 
ref.\cite{Bali-all}. 
We found that the numerical data for $E_z$ and
$k_{\theta}$ are strikingly well reproduced
by a classical solution of the model (\ref{sigma-lattice}). The discrepancy
with the observed magnetic currents distribution mentioned above disappears
completely. Fig.~2 represents the results of the fit. The corresponding
values of the parameters are:

\bea{ll}\label{param-values-lattice}
g/2\pi  & =    0.9519 \pm  0.0041 \, , \\
m_V & =  0.4522 \pm   0.0206 =  (1.0351 \pm 0.0472) \mbox{~GeV}\, , \\
m_H & =    0.4747 \pm   0.0600 =  ( 1.0866 \pm  0.1373 ) \mbox{~GeV} \,\, .
\eea
It appears that vector and scalar masses are equal to each other again
although they are now $\sim$23\% smaller than in (\ref{param-values}). Thus
the seeming discrepancy between $SU(2)$ gluodynamics and classical effective
dual AHM is recognized as merely a discretization artifact.  
The string tension (\ref{sigma-lattice}) turns out to be

\beq\label{sigma-lattice-fit}
\sqrt{\sigma} \; =  \; 0.1808 \pm 0.0213 \; \approx
\; ( 414.8 \pm 48.9 ) \mbox{~MeV} \; \approx \; 0.94
\; \sqrt\sigma_{SU(2)} \, \, .
\eeq

{\bf 4.}
To summarize, we have shown that the Abelian electric field profile and the
magnetic current distribution around the static confining string in $SU(2)$
gluodynamics may be well described as a classical ANO vortex. Moreover,
we found that the {\it classical} energy per unit length of the ANO vortex
accounts for 94 \% of the {\it quantum} $SU(2)$ string tension. The
effective dual Abelian Higgs Model appears to lie on the border
between type-I and type-II superconductivity. The magnetic charge of the
Higgs field \eq{param-values-lattice} corresponds to an electric charge of the
quark $q \approx 0.95$ (the charge of the quark in $SU(2)$ Wilson loop is
unity). This approximate correspondence of classical AHM and quantum $SU(2)$
gluodynamics 
is dynamically more specific than just Abelian dominance \cite{AbDom}.
The detailed study of this correspondence implies investigation of the
Abrikosov vortex in quantum AHM, which is in progress now.

\vspace{5mm}

We acknowledge thankfully discussions with M.~N.~Chernodub and G.~Bali.
Also we are very grateful to G.~Bali for providing us with the lattice data.
M.~I.~P. feel much obliged for the kind hospitality extended to him at the
Kanazawa University where this work has been started. The work of F.~V.~G.
and M.~I.~P. was partially supported by RFBR 99-01230a, RFBR 
-96-15-96740, INTAS 96-370 and JSPS grants.

\vfill

\subsubsection*{Figure Captions.}

\begin{itemize}
\item[Fig.~1] The fit of the data of Ref. \cite{Bali-all} by
a continuum solution.

\item[Fig.~2] The same as in Fig.~1 but using a classical solution on the
two-dimensional lattice.
\end{itemize}


\newpage
\thispagestyle{empty}

\begin{figure}[ht]
\centerline{\psfig{file=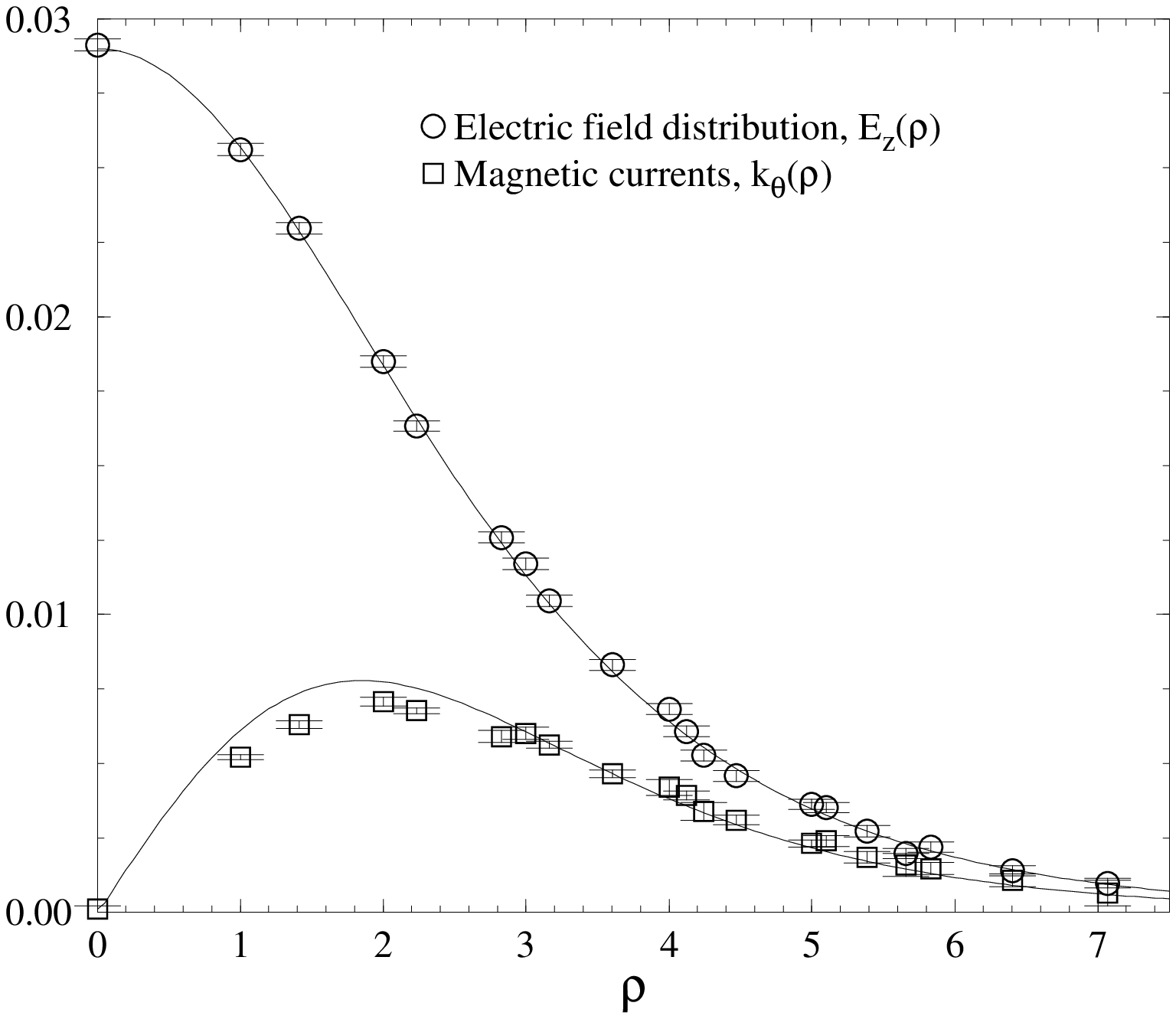,height=0.45\textheight,silent=}}
\centerline{{\bf Fig.~1}}

\vspace{7mm}

\centerline{\psfig{file=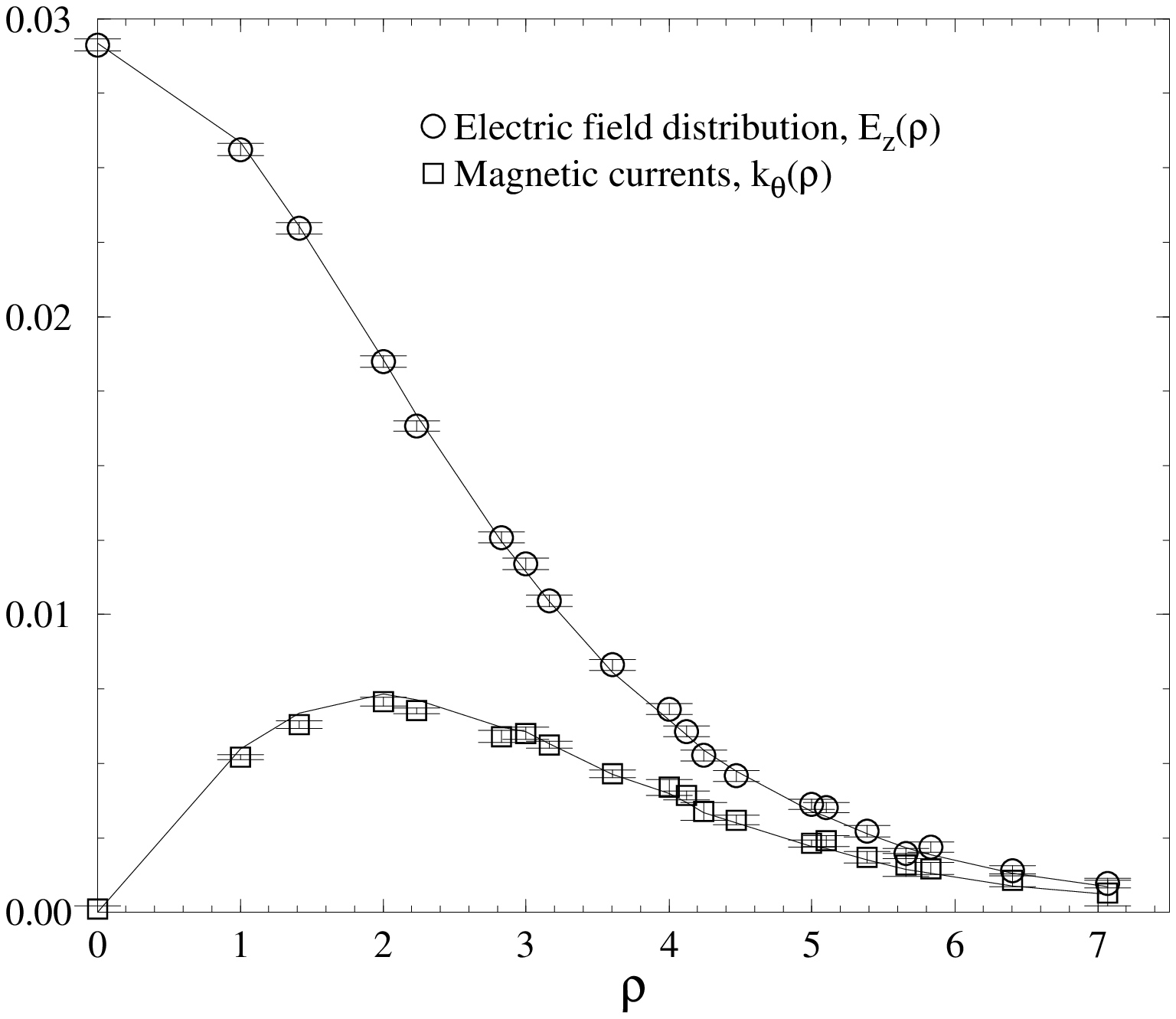,height=0.45\textheight,silent=}}
\centerline{{\bf Fig.~2}}
\end{figure}

\end{document}